\RequirePackage{fix-cm}
\documentclass[smallextended]{svjour3}       
\smartqed  
\usepackage{graphicx}
\usepackage{url}
\usepackage{natbib}
\usepackage{color}

\begin{document}

\title{\textbf{\textit{Herschel}} SPIRE FTS telescope model correction\thanks{{\it Herschel} is an ESA space observatory with science instruments provided by European-led Principal Investigator consortia and with important participation from NASA.}}

\titlerunning{{\textit{Herschel}} SPIRE FTS telescope model correction}        

\author{Rosalind Hopwood \and Trevor Fulton \and Edward T. Polehampton \and Ivan Valtchanov \and Dominique Benielli \and Peter Imhof \and Tanya Lim \and Nanyao Lu \and Nicola Marchili \and Chris P. Pearson \and Bruce M. Swinyard
}
\authorrunning{Hopwood et al.} 

\institute{R. Hopwood \at
              Imperial College London, Blackett Laboratory, Prince Consort Road, London SW7 2AZ, UK \\
              \email{rhopwood@imperial.ac.uk}           
           \and
           T. Fulton \at
              Blue Sky Spectroscopy, Lethbridge, AB, T1J0N9, Canada\\
              University of Lethbridge, Lethbridge, AB T1K3M4, Canada
              \and
              E. T. Polehampton \at
              University of Lethbridge, Lethbridge, Alberta T1K3M4, Canada\\
              Rutherford Appleton Laboratory, Chilton, Oxfordshire OX11 0QX, UK
              \and
              I. Valtchanov \at
              ESAC, P.O. Box 78, 28691 Villanueva de la Ca\~nada, Madrid, Spain
              \and
              D. Benielli \at
              Aix Marseille Universit\'e, CNRS, LAM (Laboratoire d'Astrophysique de Marseille) UMR 7326, 13388, Marseille, France
              \and 
              P. Imhof \at
              Blue Sky Spectroscopy, Lethbridge, AB, T1J0N9, Canada\\
              University of Lethbridge, Lethbridge, AB T1K3M4, Canada
              \and 
              T. Lim \at
              Rutherford Appleton Laboratory, Chilton, Oxfordshire OX11 0QX, UK
              \and 
              N. Lu \at
              NHSC/IPAC, 100-22 Caltech, Pasadena, CA 91125, USA
              \and 
              N. Marchili \at
              Universit\'a di Padova, I-35131 Padova, Italy
              \and
              C. P. Pearson \at
              Rutherford Appleton Laboratory, Chilton, Oxfordshire OX11 0QX, UK\\
              The Open University, Milton Keynes MK7 6AA, UK
              \and 
              B. M. Swinyard \at
              Rutherford Appleton Laboratory, Chilton, Oxfordshire OX11 0QX, UK \\
              University College London, Gower St, London, WC1E 6BT, UK
}

\date{Received: date / Accepted: date}

\maketitle

\begin{abstract}
Emission from the {\it Herschel} telescope is the dominant source of radiation for the majority of SPIRE Fourier transform spectrometer (FTS) observations, despite the exceptionally low emissivity of the primary and secondary mirrors. Accurate modelling and removal of the telescope contribution is, therefore, an important and challenging aspect of FTS calibration and data reduction pipeline. A dust-contaminated telescope model with time invariant mirror emissivity was adopted before the {\it Herschel} launch. However, measured FTS spectra show a clear evolution of the telescope contribution over the mission and strong need for a correction to the standard telescope model in order to reduce residual background (of up to 7\,Jy) in the final data products. Systematic changes in observations of dark sky, taken over the course of the mission, provide a measure of the evolution between observed telescope emission and the telescope model. These dark sky observations have been used to derive a time dependent correction to the telescope emissivity that reduces the systematic error in the continuum of the final FTS spectra to $\sim$0.35\,Jy.

\end{abstract}

\section{Introduction}
\label{intro}
The Spectral and Photometric Imaging REceiver \citep[SPIRE:][]{Griffin10} is one of three focal plane instruments aboard the ESA {\it Herschel} Space Observatory \citep{Pilbratt10}. SPIRE consists of an imaging photometric camera and an imaging Fourier transform spectrometer (FTS). The FTS provides simultaneous wide frequency coverage in the sub-millimetre with two bolometer arrays: SLW (447-990\,GHz) and SSW (958-1546\,GHz). As described by \citet{Griffin10}, the SPIRE instrument optical bench is cooled to roughly 4.5\,K, with the bolometric detectors at 0.3\,K. In contrast, the operational temperature of the {\it Herschel} primary mirror is 80-90\,K \citep{Pilbratt10} and so emission from the telescope dominates nearly all FTS data. Accurate removal of the telescope contribution is, therefore, an important stage of FTS data processing for all observations and crucial for observations of faint astronomical targets. Over the course of the mission, FTS spectra show an increase in background residual, after subtraction of the standard telescope model, with observations from later operational days (ODs) presenting continuum levels of as many as tens of Jy away from those predicted. 
Such significant residuals indicate the standard model fails to describe the telescope contribution sufficiently, over a long period of time.
This paper reports an investigation into the {\it Herschel} telescope emission using FTS observations of a region of sub-millimetre dark sky in order to derive a time dependent correction to the telescope model. Section \ref{sect:telMod} describes the standard telescope emission model. The data set used to derive the correction is discussed in Sect. \ref{sec:data} and the derivation and application of the correction described in Sect. \ref{sec:mod}. Results and conclusions are presented in Sect. \ref{sec:results} and Sect. \ref{sec:conclusions}.

\section{\textit{\textbf{Herschel}} telescope model}
\label{sect:telMod}
\begin{figure*}[bct!]
\begin{center}
  \includegraphics[width=0.7\textwidth]{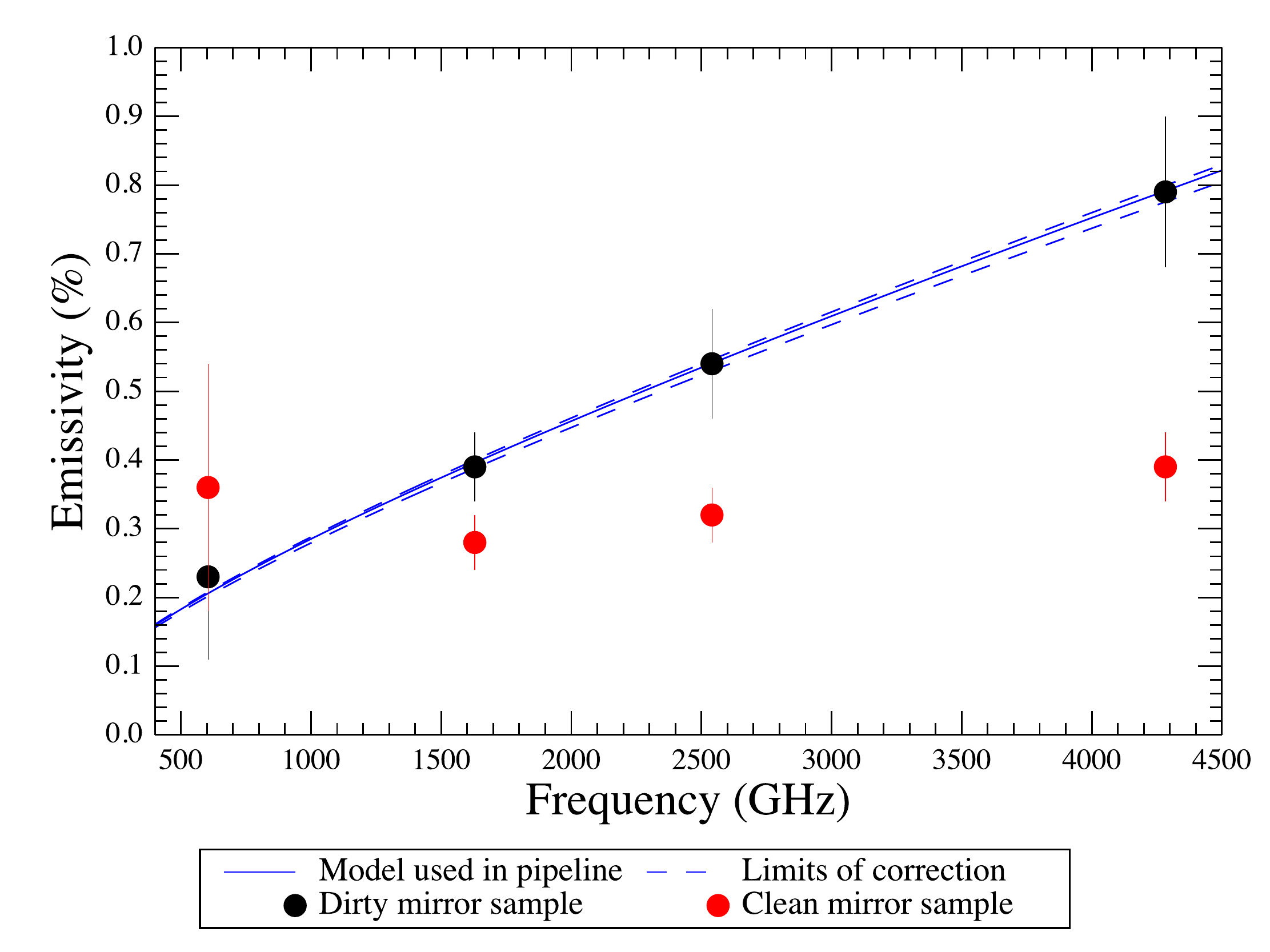}
\end{center}
\caption{Emissivity model, fitted to the dust-contaminated mirror sample by \citet{Fischer04}.}
\label{fig:emissivity}       
\end{figure*}

We assume the {\it Herschel} telescope mirrors are black body emitters, with emissivity ($\varepsilon$) determined from ground based tests by \citet{Fischer04}. The emissivity of both clean and dust-contaminated mirror samples were measured (see Fig. \ref{fig:emissivity}). The model of the telescope ($M_{\rm Tel}$) is constructed using the dust-contaminated \citet{Fischer04} emissivity best fit and two black body emitters for the primary (M1) and secondary (M2) mirrors as 

\begin{equation}
\label{equ:teleMod}
M_{\rm Tel}(\nu) = (1 - \varepsilon_{\rm M2}(\nu))\varepsilon_{\rm M1}(\nu)B(T_{\rm M1},\nu) + \varepsilon_{\rm M2}(\nu)B(T_{\rm M2},\nu),
\end{equation}

\noindent where $T_{\rm M1}$ and $T_{\rm M2}$ are the temperatures of the primary and secondary mirrors, $\nu$ is in GHz, $\varepsilon_{\rm M1}$ and $\varepsilon_{\rm M1}$ are defined following \citet{Fischer04} as


\begin{equation}
\label{equ:emissivity}
\varepsilon_{\rm M1}(\nu) = \varepsilon_{\rm M2}(\nu) = 1.0625\left(\frac{\nu}{c}\right)^{0.5} + 273.0\left(\frac{\nu}{c}\right)
\end{equation}

\noindent and $B(T,\nu)$ represents the Planck function

\begin{equation}
B(T,\nu) = \frac{2h\nu^3}{c^2}\frac{1}{e^{h\nu/kT}-1}.
\end{equation}

\begin{figure*}
\begin{center}
  \includegraphics[width=0.50\textwidth]{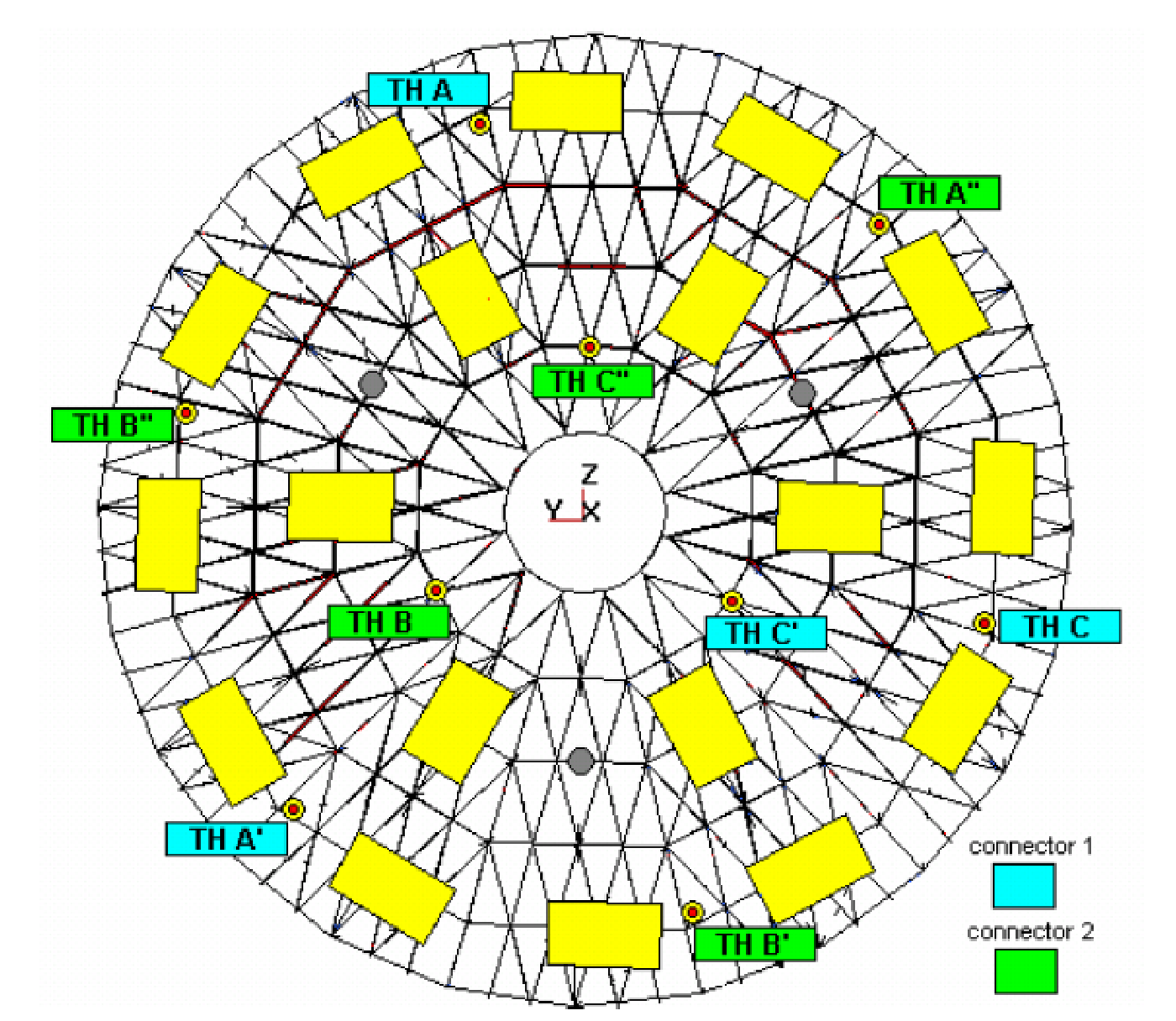}
  \includegraphics[width=0.45\textwidth]{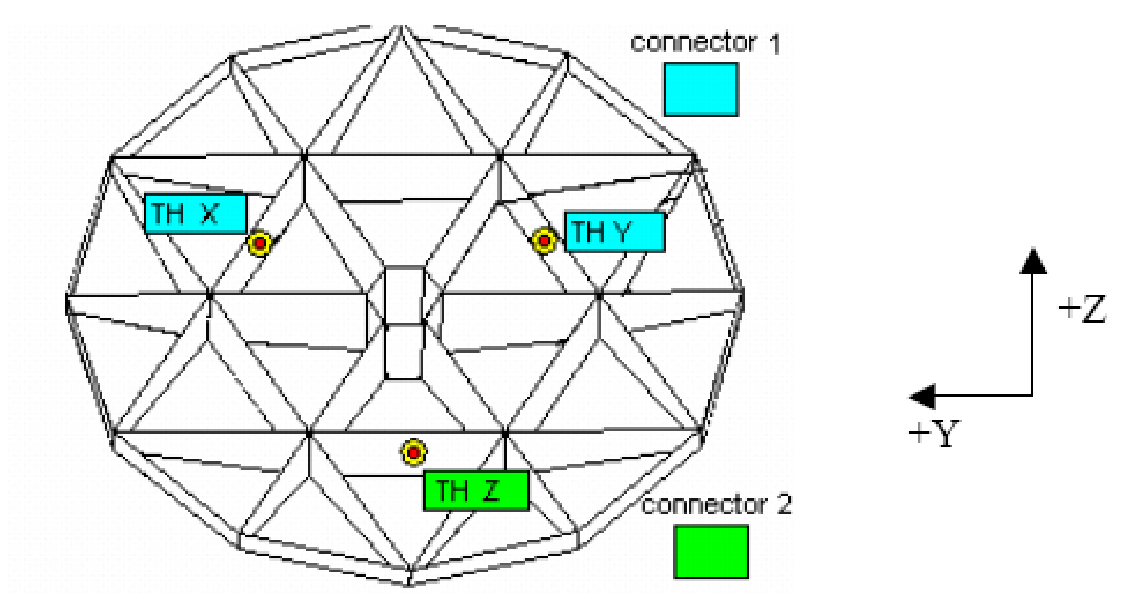}
\end{center}
\vspace*{-0.3cm}
\caption{Locations of the 9 thermistors on the primary mirror (left) and 3 thermistors on the secondary mirror (right). }
\label{fig:thermom}       
\end{figure*}

\begin{figure*}
\begin{center}
\includegraphics[width=1.0\textwidth]{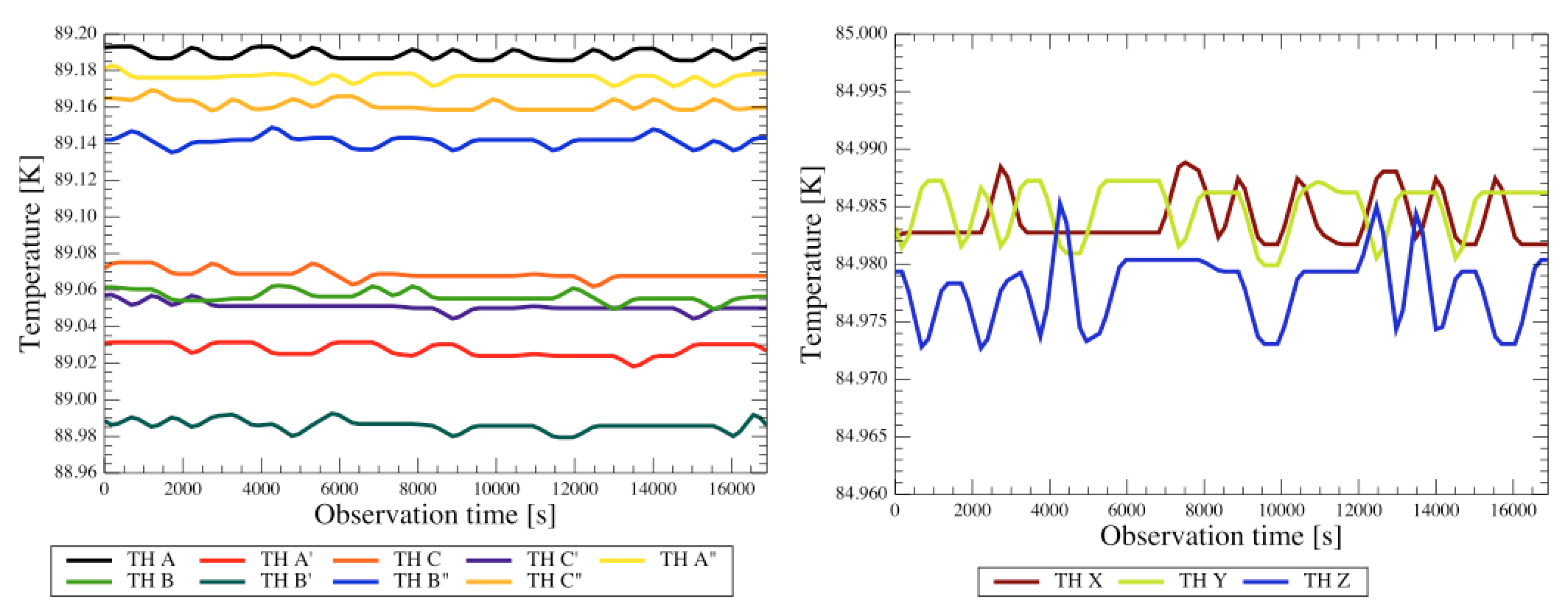}
\end{center}
\vspace*{-0.3cm}
\caption{Primary mirror (left) and secondary (right) thermometer timelines for the dark sky observation 0x50007F42 (1342209858). The recorded temperatures levels are relatively stable over the duration of the observation ($\sim$\,4.5 hours), with 1$\sigma$ errors of 0.003\,K for all thermistors. See Fig. \ref{fig:thermom} for the respective thermistor positions.}
\label{fig:termTemp}       
\end{figure*}

\noindent The temperatures of the {\it Herschel} telescope mirrors are measured by thermistors positioned on the back of each mirror, the locations of which are shown in Fig. \ref{fig:thermom}. The thermistors are read out every 512 seconds and the resulting temperatures included in the satellite telemetry for each FTS observation.
There are differences between thermometers in the overall temperature levels recorded, as shown in Fig. \ref{fig:termTemp}. As these levels are assumed to be stable over the time scale of an observation and each SPIRE detector views the whole telescope mirror, for any given observation the mean temperature is taken for each set of mirror thermometer timelines and time averaged when constructing $M_{\rm Tel}$.\\

\section{Observations}
\label{sec:data}
\begin{figure*}
\begin{center}
  \includegraphics[width=0.95\textwidth]{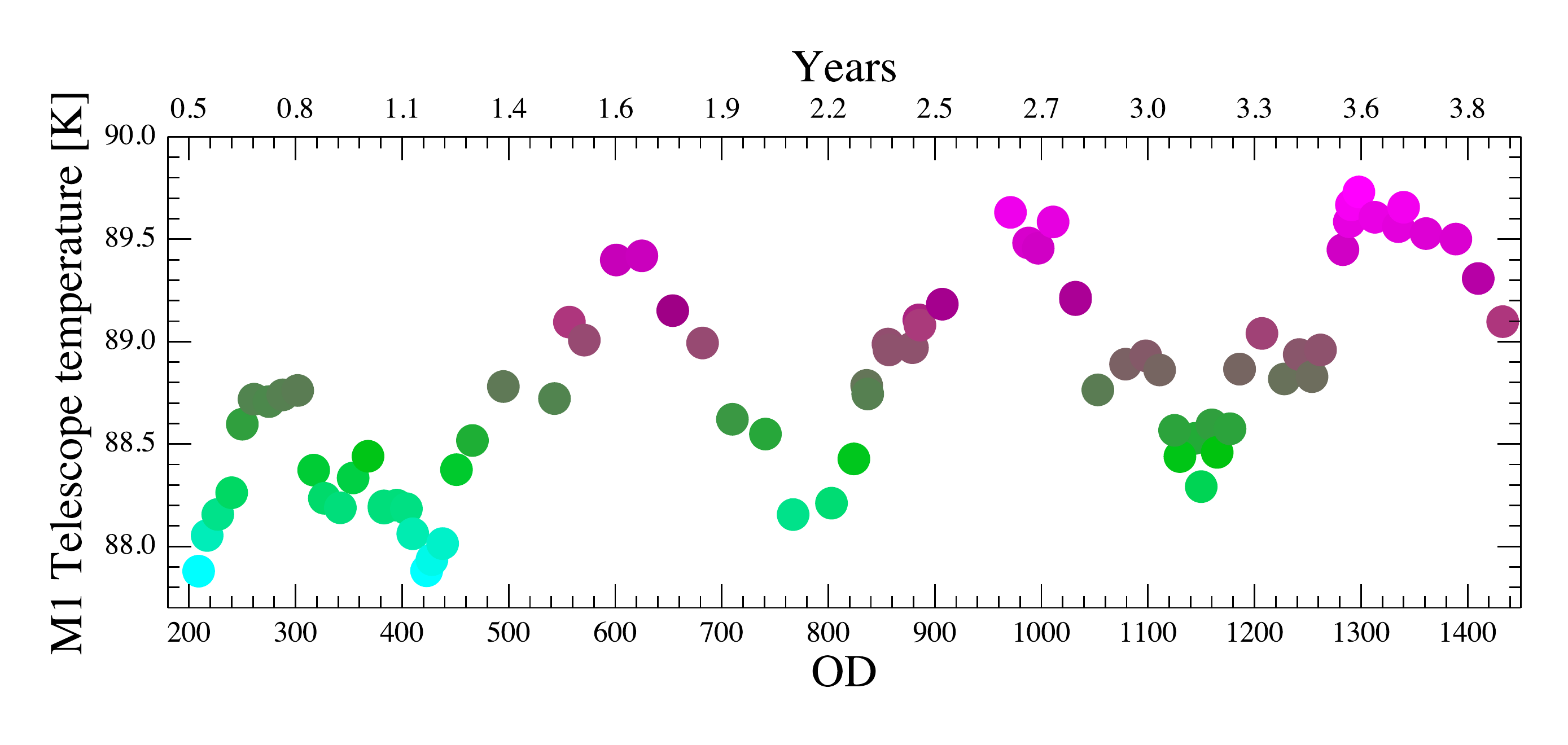}\\\includegraphics[width=0.95\textwidth]{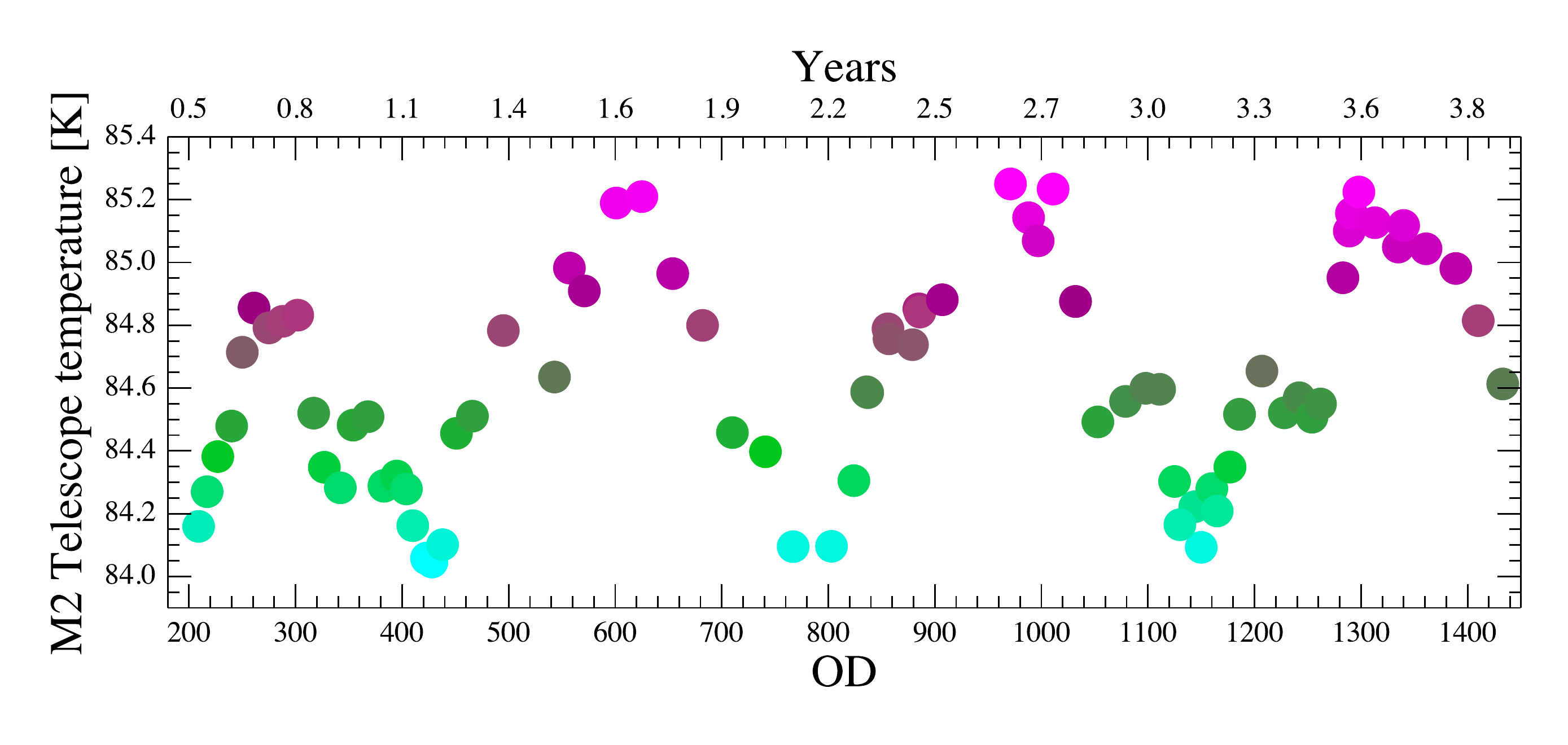}
\end{center}
\caption{Telescope mirror temperatures (top: primary mirror M1, bottom: secondary mirror M2) plotted as a function of operational day for dark sky observations. Points are coloured according to temperature. There is a sinusoidal trend in mirror temperature for both the primary and secondary mirrors, which has a period of around one year. For the secondary mirror, there is no overall change to the minimum and maximum temperatures, illustrated by the constant colours at the peaks and troughs with time. The recorded primary mirror temperatures do exhibit an overall linear rise, which is highlighted in the top plot by changing colours for the peaks and troughs of the seasonal trend.}
\label{fig:temps1}       
\end{figure*}

Observations of the nominated SPIRE dark sky field act as a useful diagnostic for FTS calibration. These dark sky observations were regularly taken over the duration of the mission, with pointing coordinates of RA:17h40m12s and Dec:+69d00m00s (J2000). High resolution dark sky observations are available for most SPIRE FTS ODs and after OD\,557 (22$^{\rm nd}$ June 2010), a dark sky at least as long as the longest science observation was taken per pair of FTS ODs. All sparse, nominal mode observations of dark sky are listed on the {\it Herschel} public TWiki\footnote{http://herschel.esac.esa.int/twiki/bin/view/Public/SpireDailyDarkObservations}. Assuming a perfect instrument and telescope subtraction and no contribution from real astronomical signal, a dark sky spectrum should purely consist of random noise distributed about zero. Any longterm trend away from this expected null signal,
indicates a systematic divergence from the standard telescope model. Such trends, however, can be used to derive a time dependent empirical correction. The first indication of a change in observing conditions with time can be seen in Fig. \ref{fig:temps1}, which compares primary and secondary mirror temperatures for all long ($>\,20$ repetitions) dark sky observations. There is a clear periodic trend that cycles over the scale of a year and is similar for both sets of temperatures, but the primary mirror temperature also exhibits an overall increase with time, whereas the secondary mirror temperature presents no such trend. The overall increase in $T_{\rm M1}$ implies a gradual change in observing conditions for the primary mirror during {\it Herschel} operations (e.g. due to a buildup of dust), which is not reflected in the telescope model by the time invariant emissivity term.

\section{Correcting the telescope model}
\label{sec:mod}
When reducing FTS data (see \citealt{Fulton13b} for details on the standard FTS pipeline), the telescope model is used in two places: firstly, for deriving the telescope relative spectral response function ($R_{\rm Tel}$;  \citealt{Swinyard13}, \citealt{Fulton13a})
and secondly, to subtract the emission of the telescope from the extended calibrated spectrum. The measured voltage density spectrum (in units of V\,GHz$^{-1}$) of an FTS observation can be expressed as the linear combination of contributions from the astronomical source ($V_{\rm source}$), the telescope ($V_{\rm Tel}$) and the SPIRE instrument ($V_{\rm Inst}$). For a dark sky observation $i$, the measured voltage density spectrum ($V_i$) contains only $V_{\rm Tel}$ and $V_{\rm Inst}$ giving

\begin{equation}
\label{eq:vmeas1}
V_i(\nu) = {V_{\rm Tel}(\nu)} + {V_{\rm Inst}(\nu)},
\end{equation}

\noindent where ${V_{\rm Tel}(\nu)}$ = $M_{\rm Tel}(\nu)R_{\rm Tel}(\nu)$.\\

\noindent After accounting for the instrument emission, Eq. \ref{eq:vmeas1} reduces to

\begin{equation}
\label{eq:vmeas2}
V_{i}^\prime(\nu) = {V_{\rm Tel}(\nu)},
\end{equation}

\noindent and at this stage, any dark sky observation can be used to calculate a daily telescope RSRF $R_{\rm Tel}^i$ by dividing $V_{i}^\prime$ by the respective $M_{\rm Tel}$. A ratio of daily to generic telescope RSRF ($\overline{R}_{\rm Tel}$) can then be taken, where $\overline{R}_{\rm Tel}
$ is derived from many dark sky observations (see Fulton et al. 2013 for full details of the derivation) and taken from the SPIRE FTS calibration tree. $\overline{R}_{\rm Tel}$ is compared to $R_{\rm Tel}^i$ for many dark sky observations to derive a correction factor to the standard telescope model.
The comparison is performed by taking ratios r, found per observation $i$, as

\begin{equation}
\label{equ:ratio}
r_i(\nu) = \frac{1}{\overline{R}_{\rm Tel}(\nu)}\frac{V_i(\nu)}{M_{\rm Tel}(\nu)}.
\end{equation}

\noindent The median of $r_i(\nu)$ is taken to give a single value $\widetilde{r}_i$, per observation.\\

\begin{figure*}[b!]
\begin{center}
  \includegraphics[width=0.95\textwidth]{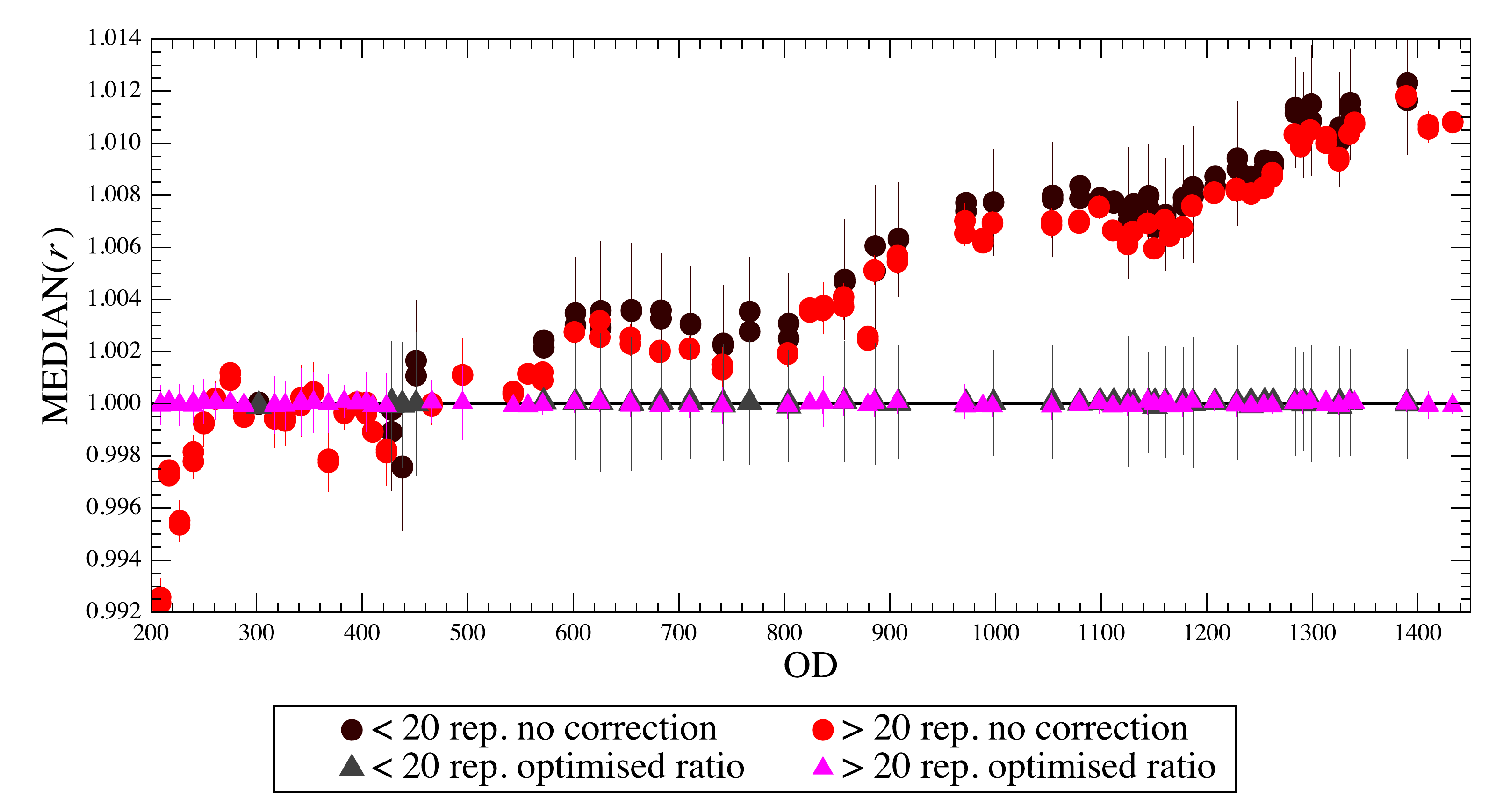}
\end{center}
\vspace*{-0.3cm}
\caption{Median ratios (i.e. $\widetilde{r}_i$) of daily to generic telescope RSRF ($R_{\rm Tel}^i$ and $\overline{R}_{\rm Tel}$) for the SLW centre detector (SLWC3) as a function of OD. Circles: ratios taken without applying the telescope model correction. Triangles: the ratios optimised to 1.00 following method described in the main text. Red and magenta symbols show the ratios for observations with $>$\,20 repetitions and black and dark grey symbols indicate ratios for observations with $<$\,20 repetitions. There is a strong diversion away from the ideal value of 1.0 for ratios without any correction applied to the telescope model. Errors are assigned from the standard deviation of the respective $r_i(\nu)$. The $<$\,20 repetition observations consist mainly of 5 repetition dark sky. These darks were nearly always taken at the end of an FTS observing cycle (two concatenate ODs), which means a higher than average observing temperature than other observations taken the same cycle. This difference in temperature translates to a slight systematic shift in their ratios.}
\label{fig:ratios1}       
\end{figure*}

\begin{figure*}[bct!]
\begin{center}
  \includegraphics[width=0.95\textwidth]{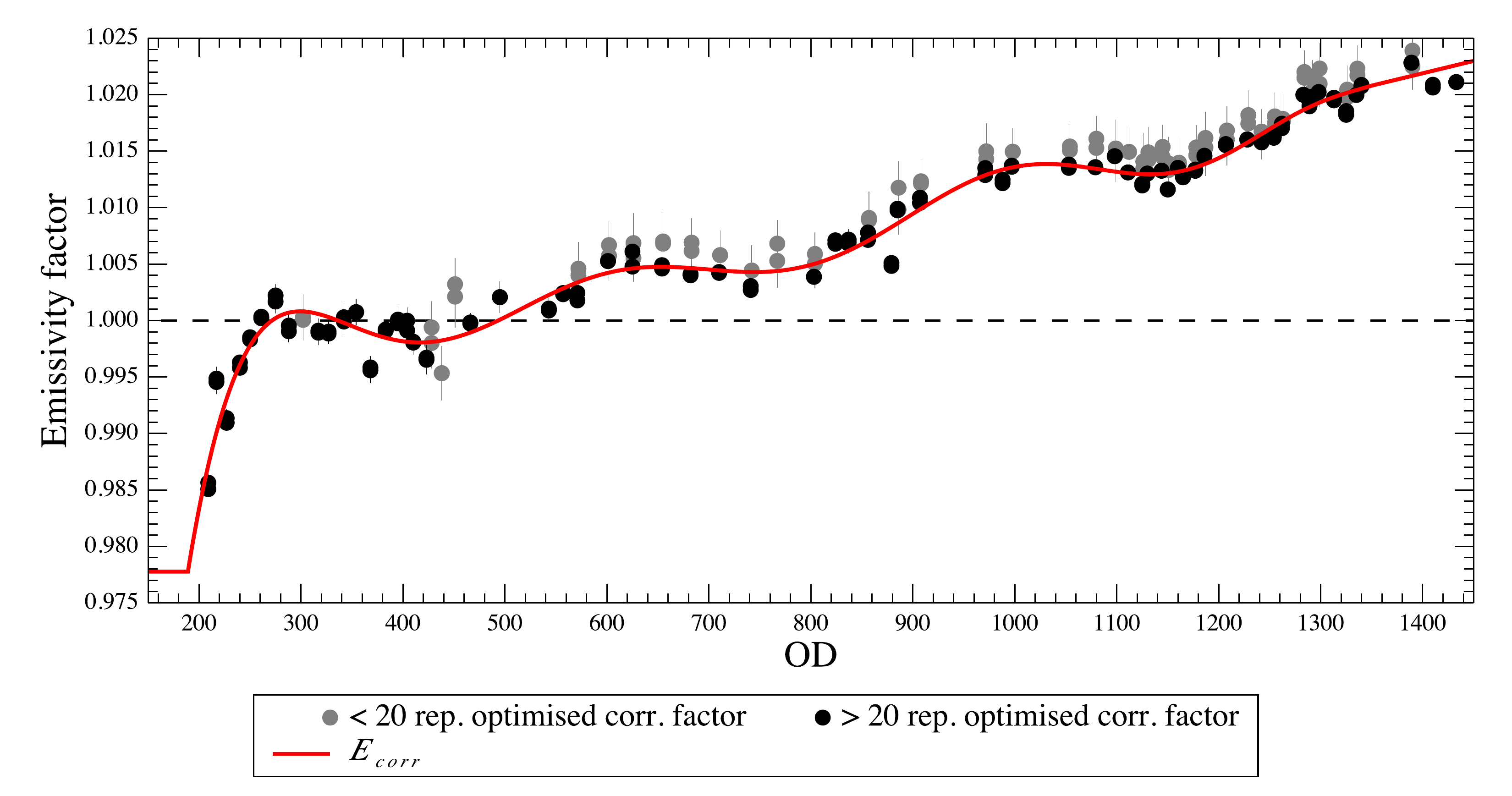}
\end{center}
\vspace*{-0.3cm}
\caption{Optimised telescope model corrections for the centre detectors SLWC3 and SSWD4 and corresponding final correction. The black (observations with $>$\,20 repetitions) and grey (observations with $<$\,20 repetitions) points are the correction factors that correspond to the optimised ratios in Fig. \ref{fig:ratios1} (magenta and grey triangles, respectively) with the standard deviation of the corresponding $r_i(\nu)$ assigned as errors. Both a linear and a polynomial fit are performed to the corrections and the best fits are used to construct the final correction (red line), see main text for further details.}
\label{fig:fit1}       
\end{figure*}

\begin{figure*}[bct!]
\begin{center}
  \includegraphics[width=0.49\textwidth]{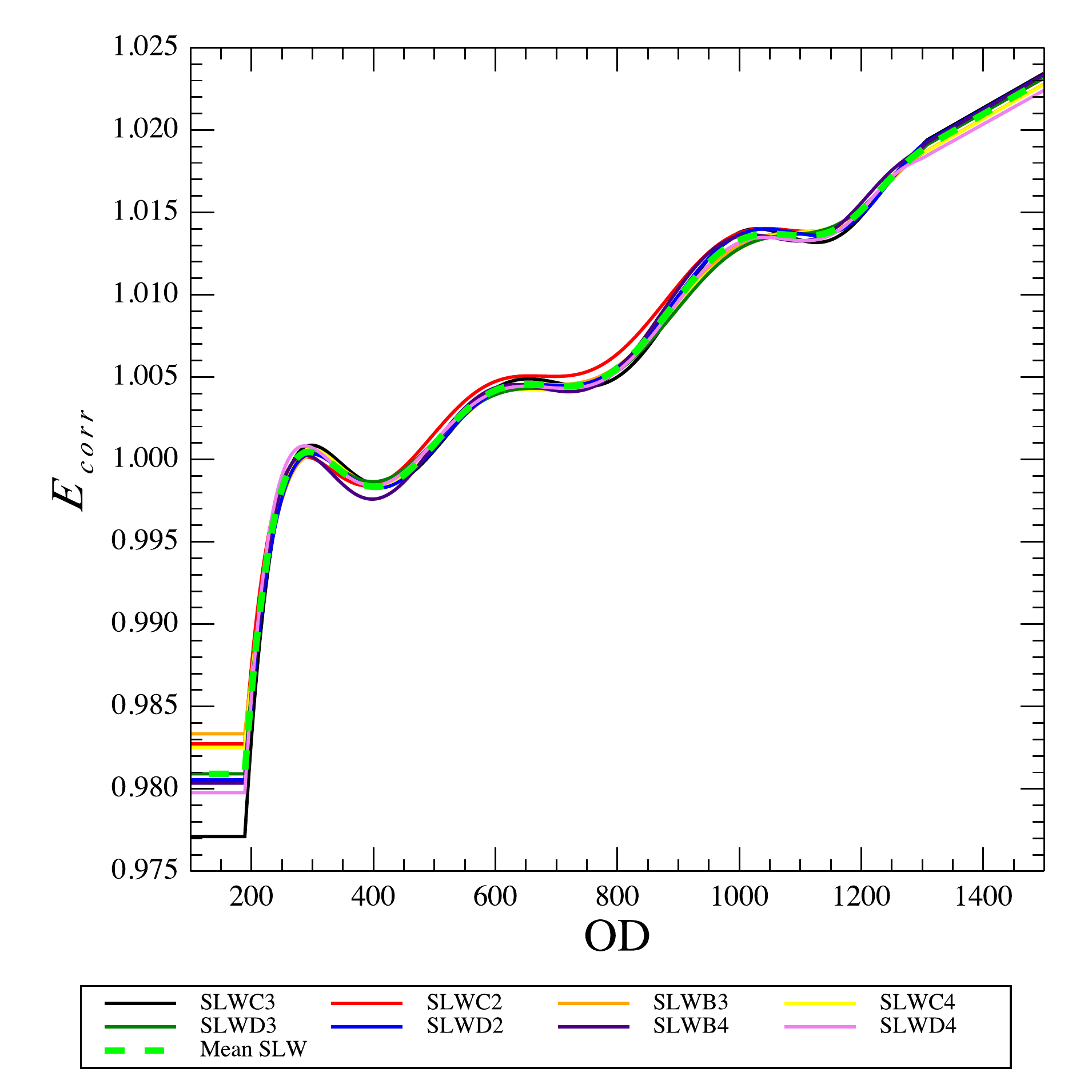}
  \includegraphics[width=0.49\textwidth]{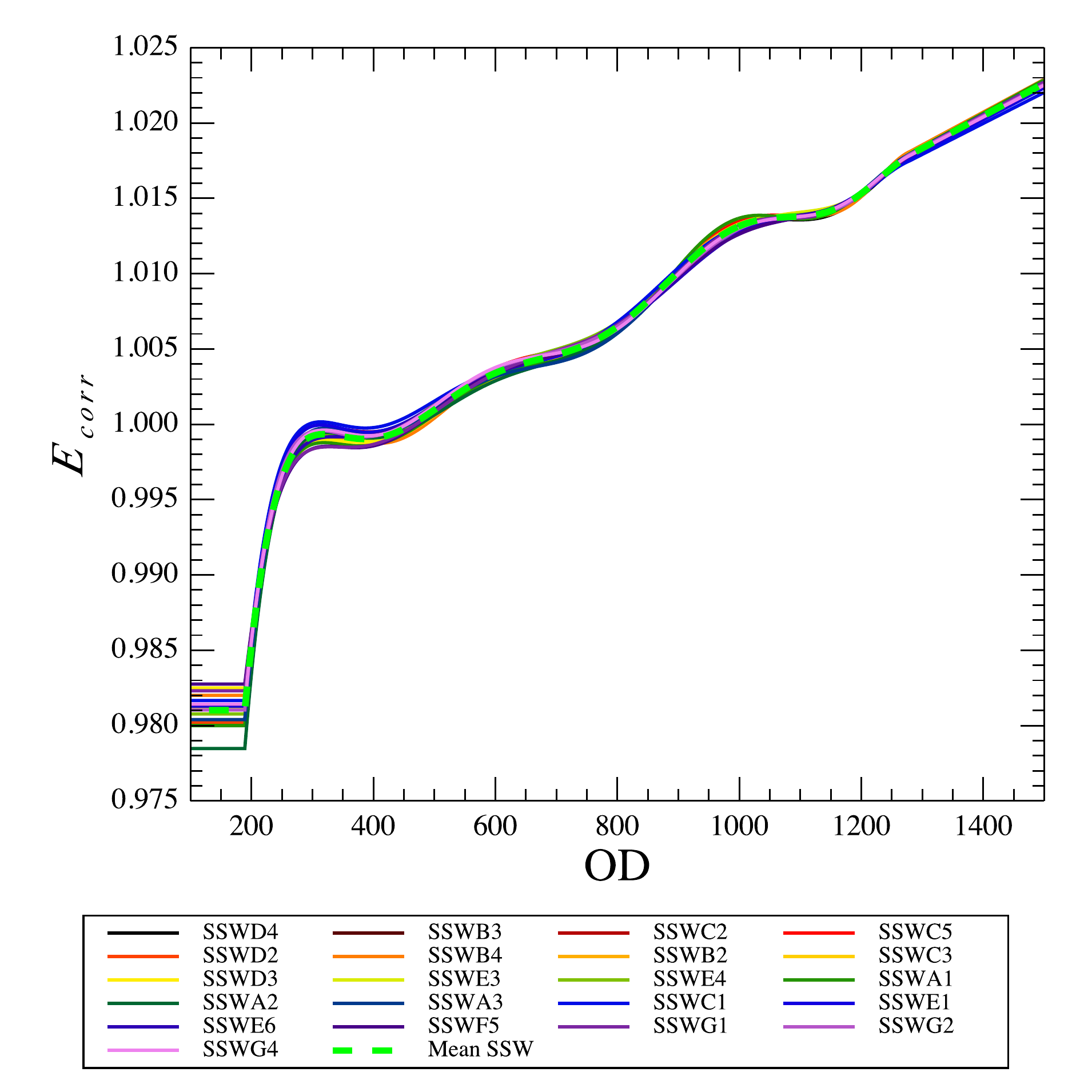}
\end{center}
\vspace*{-0.1cm}
\caption{Mean telescope model corrections (dashed green) for SLW (left) and SSW (right) shown over all the individual detector corrections used. The detectors for which the mean correction is used are: for SLW -- SLWC5, SLWA1, SLWB1, SLWC1, SLWA3, SLWD1 and SLWE1; for SSW -- SSWF3, SSWF2, SSWE2, SSWD7, SSWD6, SSWC6, SSWG3, SSWF1, SSWD1, SSWB5 and SSWC4.}
\label{fig:meancorr}       
\end{figure*}

\noindent To optimise the correction to the telescope model for any given detector the following steps are taken: 
\begin{itemize}
\item A set of suitable observations is defined on a detector by detector basis. Primarily observations of the SPIRE dark sky field are used, however observations of extremely faint targets are included for some detectors to extend temporal coverage. In addition, observations of point sources can be included for off-axis detectors, however only after careful inspection for nearby source or extended background contamination. 
\item $\widetilde{r}_i$ is determined for each observation $i$, in the selected set, using Eq. \ref{equ:ratio} (see Fig. \ref{fig:ratios1} for the resulting ratios plotted for one detector).
\item $\widetilde{r}_i$ is calculated with a multiplicative correction factor applied to the M1 emissivity of $M_{Tel}$. This correction factor is then adjusted by an arbitrary step size to provide an $\widetilde{r}_i$ closer to 1.0. This process is repeated until $\widetilde{r}_i$ is within 10$^{-4}$ of 1.0. The optimised correction factors for SLWC3 are plotted in Fig. \ref{fig:fit1} as black and grey filled circles. 
\item A fit, weighted by the associated uncertainties, is performed to the resulting optimised correction factors with a high order polynomial, and a separate linear function.
\item The final model correction (see Fig. \ref{fig:fit1}) is constructed as a function of OD by combining:
    \begin{itemize}
    \item the fitted polynomial at OD189, fixed over the performance verification phase of the mission (ODs 0-189);
    \item the fitted polynomial between ODs 189-1280 for SSW and ODs 189-1308 for SLW; 
    \item and, due to limited OD coverage, the best linear fit for days above the polynomial OD limit (OD1280 for SSW and OD1308 for SLW), after normalising to the final polynomial value used.
    \end{itemize}
\item There are several noisy detectors that are not reliably fitted due to high scatter in the associated $\widetilde{r}_i$ values. A mean correction is used for these detectors (see Fig. \ref{fig:meancorr}).
\end{itemize}


\noindent From Fig. \ref{fig:temps1}, we assume only the telescope's primary mirror emissivity ($\varepsilon_{\rm M1}$) has seen significant change over the {\it Herschel} mission and, therefore, a time dependent correction ($E_{\rm corr}$) is derived for and applied as a multiplicative factor to $\varepsilon_{\rm M1}$ in $M_{\rm Tel}$. Including this correction factor in Eq. \ref{equ:teleMod} gives

\begin{equation}
M_{\rm Tel}(\nu) = (1 - \varepsilon_{\rm M2}(\nu))\varepsilon_{\rm M1}(\nu)E_{\rm corr}B(T_{\rm M1},\nu) + \varepsilon_{\rm M2}(\nu)B(T_{\rm M2},\nu), 
\end{equation}

\noindent where $E_{\rm corr}$ is a constant that depends on the OD of a given observation.

\begin{figure*}[bct!]
\begin{center}
  \includegraphics[width=0.75\textwidth]{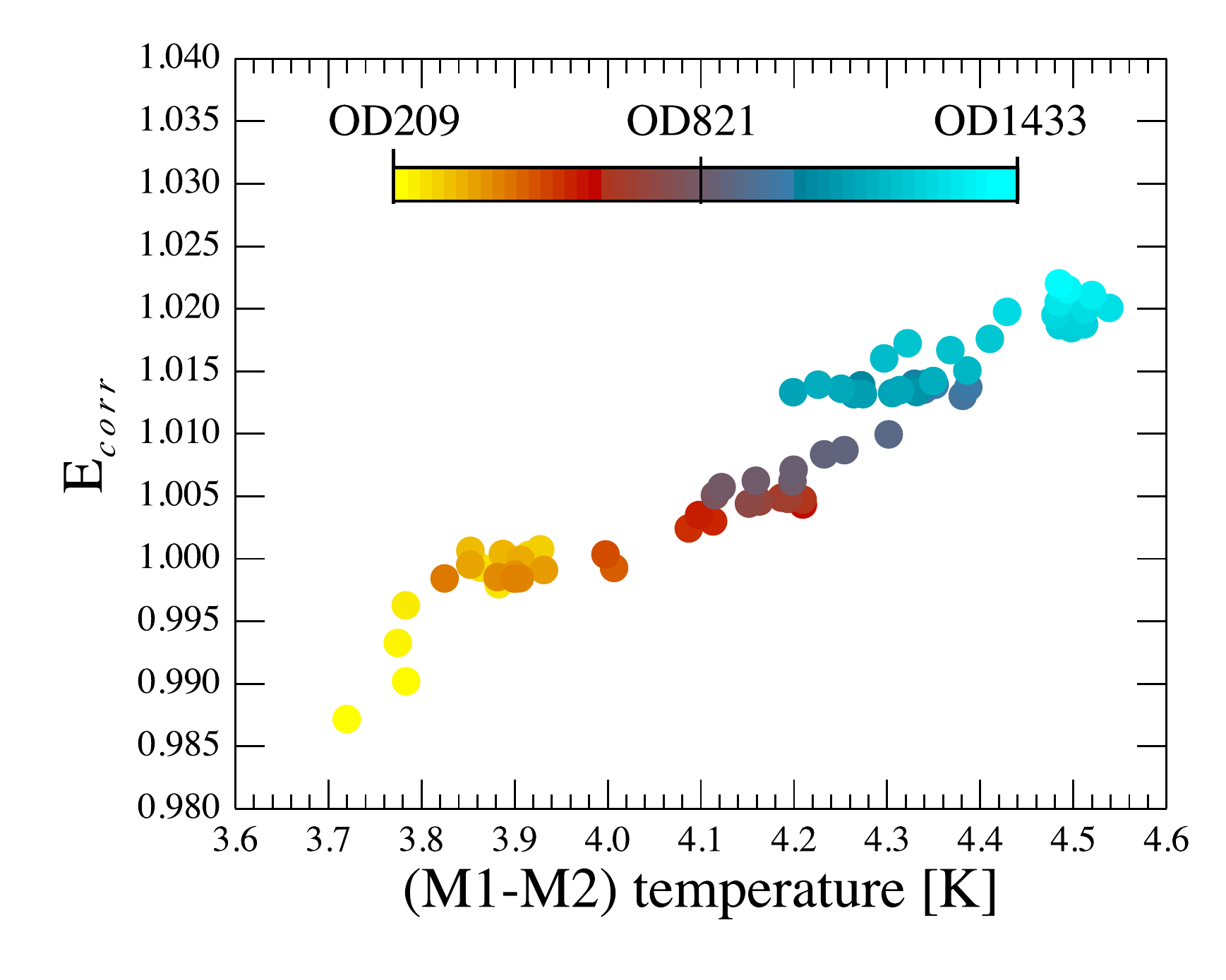}
\end{center}
\vspace*{-0.1cm}
\caption{$E_{\rm corr}$ plotted as a function of (M1-M2) telescope temperature and coloured by OD. A linear correlation can be seen, due to the increase in M1 temperature relative to M2, as indicated by Fig. \ref{fig:temps1}.
}
\label{fig:ecorrTempOd}       
\end{figure*}

\noindent The dashed blue lines in Figure \ref{fig:emissivity} show that the change seen in emissivity due to $E_{\rm corr}$ is $<$\,1\%. There is a linear correlation between the long term rise in M1 temperature (see Fig. \ref{fig:temps1}) and $E_{\rm corr}$, as shown in Fig. \ref{fig:ecorrTempOd}.

\section{Results}
\label{sec:results}
\begin{figure*}
  \includegraphics[width=0.5\textwidth]{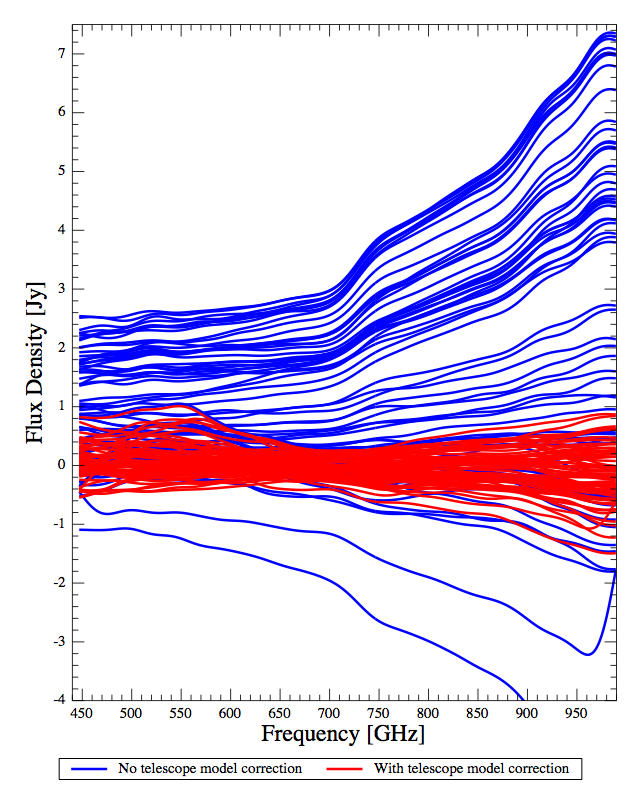}
  \includegraphics[width=0.5\textwidth]{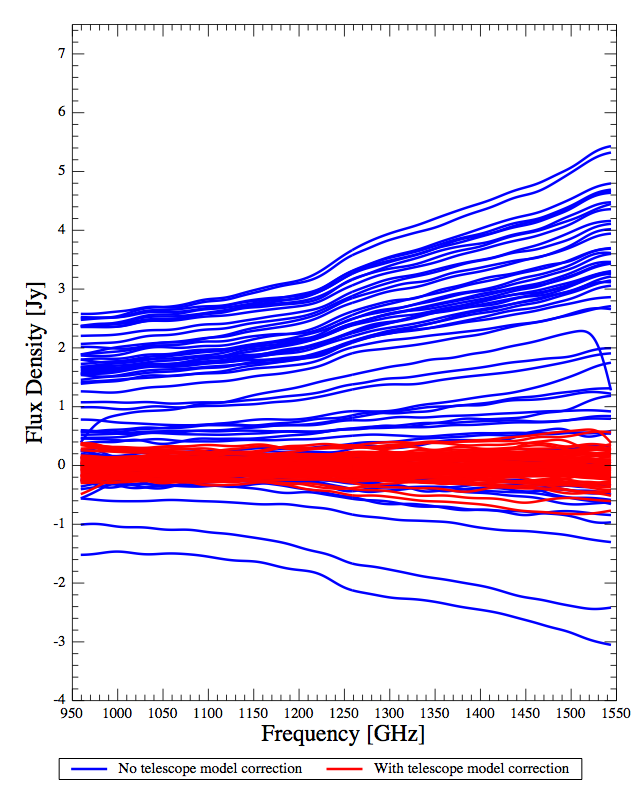}
\caption{Dark sky observations with (red) and without (blue) the empirical telescope model correction applied. All data were reduced with the standard pipeline, and smoothed to obtain the low frequency shape. There is a maximum reduction in outlying flux density of $\sim$7\,Jy for SLWC3 (left) and $\sim$5.5\,Jy for SSWD4. The corrected spectra are tightly grouped around zero, as expected for dark sky observations after an appropriate telescope subtraction.}
\label{fig:scans1}       
\end{figure*}

\begin{figure*}
\begin{center}
  \includegraphics[width=0.95\textwidth]{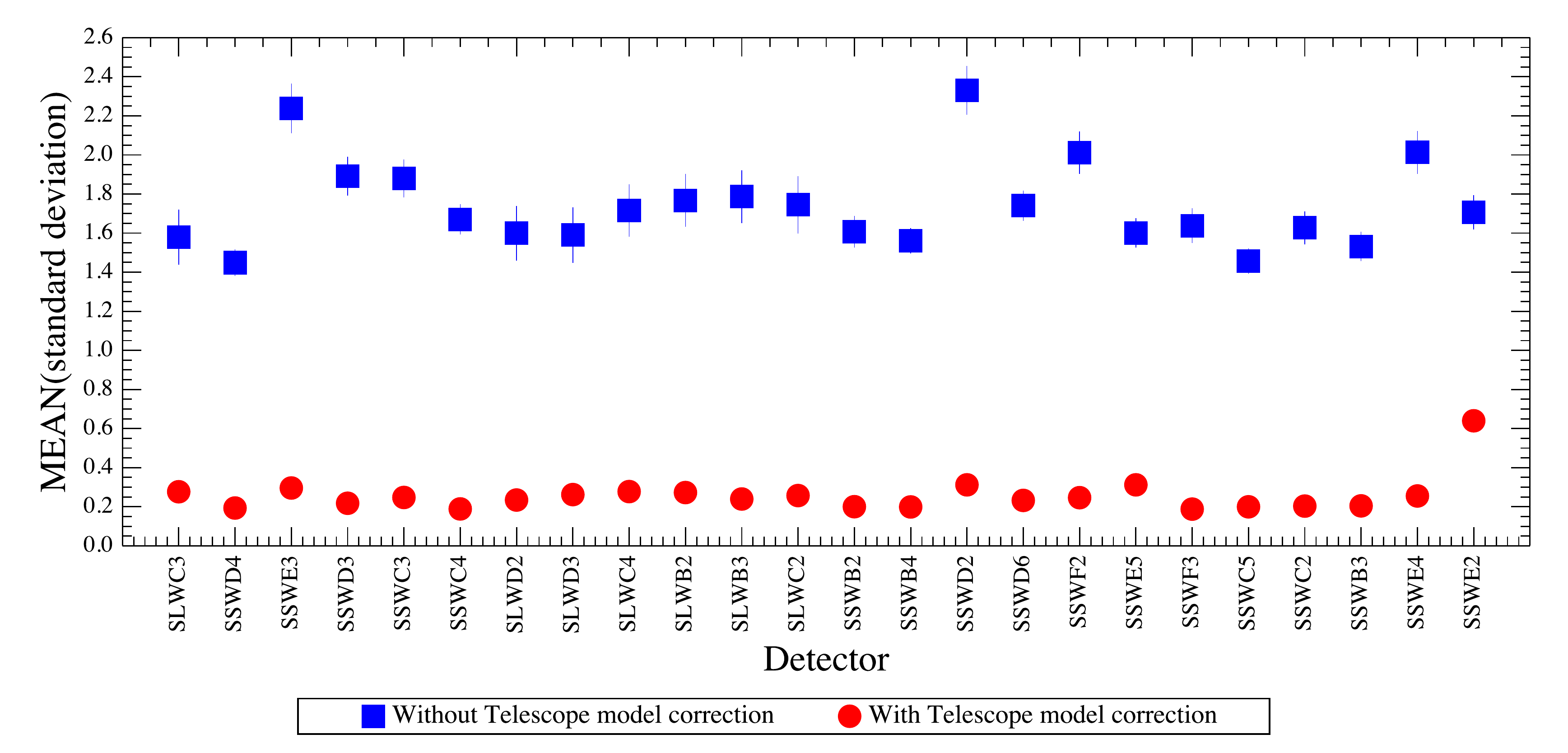}
\end{center}
\caption{For each unvignetted detector the low frequency shape was obtained for all $>$\,20 repetition dark sky observations (as shown for the centre detectors in Fig. \ref{fig:scans1}) and the standard deviation was taken to provide the spread of the low frequency shape. The mean standard deviation per detector is plotted and the results show the continuum offset error is improved for all detectors and there is more consistency between the detectors. }
\label{fig:scans2}       
\end{figure*}

In Fig. \ref{fig:scans1}, the telescope model correction is applied to the centre detectors for point source calibrated (standard pipeline level-2 products; see \citet{Fulton13b}) long ($>$\,20 repetitions) dark sky observations and compared to the uncorrected results. A significant reduction in spread is seen for the corrected level-2 data, due to the improved telescope subtraction. For SLWC3 the telescope residual is reduced by up to 7\,Jy and up to 5.5\,Jy for SSWD4. The average spread for all unvignetted detectors, across the long dark sky observation set, is shown in Fig. \ref{fig:scans2}, and illustrates the improvement seen for all detectors due to the telescope model correction.

\section{Conclusions}
\label{sec:conclusions}
An empirical correction to the telescope emissivity has been derived to compensate for changing conditions of the Herschel telescope over the course of the mission. 
Although this correction only adjusts the emissivity by $<$1\%, it is essential for FTS data to avoid significant background residual in reduced spectra. Applying the correction to data across the mission gives an improvement on the continuum of up to 7\,Jy for the centre detectors, with a mean error on the continuum of 0.4\,Jy for SLWC3 and 0.3\,Jy for SSWD4. A significantly improved background subtraction is seen for all detectors, reducing the average continuum offset error by at least 1\,Jy. 


%
\clearpage
\newpage
\begin{acknowledgements}
\textit{Herschel} is an ESA space observatory with science instruments provided by European-led Principal Investigator consortia and with important participation from NASA. SPIRE has been developed by a consortium of institutes led by Cardiff University (UK) and including Univ. Lethbridge (Canada); NAOC (China); CEA, LAM (France); IFSI, Univ. Padua (Italy); IAC (Spain); Stockholm Observatory (Sweden); Imperial College London, RAL, UCL-MSSL, UKATC, Univ. Sussex (UK); and Caltech, JPL, NHSC, Univ. Colorado (USA). This development has been supported by national funding agencies: CSA (Canada); NAOC (China); CEA, CNES, CNRS (France); ASI (Italy); MCINN (Spain); SNSB (Sweden); STFC (UK); and NASA (USA).
\end{acknowledgements}
%
\bibliographystyle{spbasic}      
\bibliography{bib1}   

\end{document}